# Renormalisation scale uncertainty in the DIS 2+1 jet cross-section

G. Ingelman[a,b] and J. Rathsman[a]

[a] Dept. of Radiation Sciences, Uppsala University, Box 535, S-751 21 Uppsala, Sweden
[b] Deutsches Elektronen-Synchrotron DESY, Notkestrasse 85, D-22603 Hamburg, FRG

**Abstract:** The Deep Inelastic Scattering 2+1 jet cross-section is a useful observable for precision tests of QCD, *e.g.* measuring the strong coupling constant $\alpha_s$. A consistent analysis requires a good understanding of the theoretical uncertainties and one of the fundamental ones in QCD is due to the renormalisation scheme and scale ambiguity. Different methods, which have been proposed to resolve the scale ambiguity, are applied to the 2+1 jet cross-section and the uncertainty is estimated. It is shown that the uncertainty can be made smaller by choosing the jet definition in a suitable way.

## 1 Introduction

Deep inelastic scattering (DIS) [1] has since long been a way of probing the nucleon structure and studying the theory of strong interactions, Quantum Chromo Dynamics (QCD). The hadronic final state has been studied at fixed target lepton scattering experiments where effects such as high-$p_\perp$ particle production [2] and the onset of jet production [3] have been observed, but the energy has not been high enough to have a substantial rate of clear multi-jet events. With the order of magnitude increased cms energy in the HERA electron-proton collider at DESY (Hamburg) the study of these phenomena enter a new era.

In fact, multijet events have already been observed at HERA [4] and with increasing statistics they will be useful for precision test of QCD. In particular, the inclusive rate of 2+1 jet events (where +1 denotes the jet emerging from the proton remnant) can be used to extract the gluon density $xg(x)$ in the proton [5] or to measure the strong coupling constant $\alpha_s$ and observe its 'running' [6], *i.e.* $\alpha_s = \alpha_s(\mu_R^2)$ where $\mu_R^2$ is the renormalisation scale. An advantage at HERA is that this running can be observed in a single experiment to avoid systematic uncertainties that may enter when combining measurements from different experiments, as has been necessary so far (see *e.g.* [7]).

In order to perform such analyses the theoretical cross-section has to be precisely defined and calculated with a proper understanding of its uncertainty. This requires the complete next-to-leading order (NLO) cross-section formalism, which also facilitates a well-defined meaning of $\alpha_s$. As with all field theories, QCD has to be renormalised to get finite answers for the predictions of physical observables. To perform the renormalisation one has to define a renormalisation scheme which defines how much of the finite parts are to be kept at each order in the perturbative expansion of an observable and a renormalisation scale which defines at which point or momentum scale the subtraction of the infinite parts should be made. From a physical point of view the result should be independent of the renormalisation scheme and scale, the so called renormalisation group invariance. This independence will, however, only hold for the complete perturbative series, whereas a truncation to any given order introduces a dependence on both the renormalisation scheme and scale (in next-to-leading order a change in the renormalisation scheme is equivalent to a change in the renormalisation scale).

To solve this ambiguity at least three different methods for choosing the renormalisation scale have been proposed, the BLM, FAC and PMS scales.

BLM or Automatic Scale Fixing as the inventors Brodsky, Lepage, Mackenzie [8] call it. The renormalisation scale is chosen so that the second order terms with $N_f$ dependence ($N_f$=number of active quark flavors) are absorbed into the running of $\alpha_s$.

FAC Fastest Apparent Convergence as proposed by Grunberg [9] is defined by requiring all second order contributions to be absorbed into the running of $\alpha_s$ so that the NLO result is equal to the LO result.

PMS Principle of Minimum Sensitivity as proposed by Stevenson [10]. The renormalisation scale is chosen so that the scale dependence is minimal, *i.e.* $\partial R/\partial \mu_R^2 = 0$ where $R$ is the physical observable under consideration.

It is the purpose of this paper to apply these alternative scale choosing methods to DIS and, in particular, use them in order to asses the theoretical uncertainty represented by this free scale. This is necessary in order to estimate the theoretical error in, *e.g.*, a measurement of $\alpha_s$ or $\Lambda_{QCD}$. Previously, only simple estimates of the uncertainty has been made by varying the renormalisation scale with arbitrary factors around the simple choice of $Q^2$ which has no theoretical justification. Here, we perform a dedicated study where the theoretical uncertainty is more properly obtained by comparing the results obtained with these better motivated scale choices.

In addition to the renormalisation ambiguities one has similar problems associated with the factorisation, which defines the way to absorb soft and collinear initial state singularities into the structure functions. This causes a corresponding scheme and scale ($\mu_F^2$) dependence for the truncated perturbative series. The PMS method has also been applied to deal with these ambiguities, originally by Politzer [11] and later for example by Aurenche et al [12] who applied the PMS method to set the factorisation scale as well as the renormalisation scale for large $p_\perp$ processes involving real photons. In this dedicated study of the renormalisation scale no attempt will be made to optimise the factorisation scale, instead there will only be some comments on the applicability of the other methods to resolve that ambiguity.

The paper is organised as follows. In section 2 we discuss the complete $\mathcal{O}(\alpha_s^2)$ DIS 2+1 jet cross-section and in section 3 we treat theoretical aspects of different renormalisation scale choices and comment on the factorisation scale dependence. The numerical results on scales and cross-sections are presented and discussed in section 4 and finally we summarise and conclude in section 5.

# 2 DIS 2+1 jet production

## 2.1 Kinematic variables and jet definition

The process of interest is (with four-momenta)

$$electron(k) + proton(P) \rightarrow electron(k') + jet1(p_1) + jet2(p_2) + remnant(p_r)$$

as illustrated in Fig. 1.

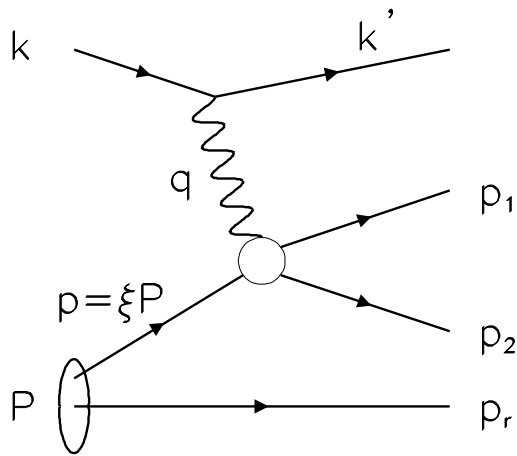

Figure 1: *Generic diagram for 2+1 jet production in DIS*

The overall *ep* cms energy squared is

$$s = (P+k)^2$$

and the inclusive deep inelastic scattering can be expressed in the variables

$$Q^2 = -q^2 = -(k-k')^2$$
$$W^2 = (P+q)^2$$

giving the squared momentum transfer and invariant mass of the hadronic final state, or alternatively by the scaling variables

$$x = \frac{Q^2}{2Pq}$$
$$y = \frac{Pq}{Pk}$$

The scaling variables

$$z_1 = \frac{Pp_1}{Pq}$$
$$z_2 = \frac{Pp_2}{Pq}$$

describe the relative energy sharing of the two jets from the hard scattering and

$$\xi = x\,(1 + \frac{(p_1 + p_2)^2}{Q^2})$$

is the incoming parton momentum fraction with respect to the proton ($p = \xi P$). In inclusive DIS, *i.e.* disregarding the hadronic final state, there are only two independent variables. These are usually taken as $x, Q^2$ or $x, y$ and they are fully determined by the energy and angle of the scattered lepton. To describe the internal degrees of freedom in the 2+1 jet system, three additional variables are needed. These are usually taken as $x_p \equiv Q^2/2pq = x/\xi$, $z = z_1$ (giving $z_2 = 1 - z$) and the azimuthal angle $\phi$ between the lepton scattering plane ($\vec{k}, \vec{k}'$) and the parton plane ($\vec{p}, \vec{p_1}$).

Experimentally, jets can be defined according to the JADE-algorithm proceeding as follows. First the invariant mass of all pairs of particles is calculated. Then the pair with smallest invariant mass is identified and if its mass is below some cut-off mass the four momenta of these two particles are added to form a pseudoparticle (according to one of several possible schemes [7]). The process is repeated until all pairs of particles or pseudoparticles have invariant masses above the cut-off and the remaining particles or pseudoparticles are then identified as jets. The cut-off is given by

$$(p_i + p_j)^2 = s_{ij} \geq y_{cut} W^2 \qquad (1)$$

where $y_{cut}$ is a parameter between zero and unity that defines the jet resolution. Since the proton remnant ($p_r$) is counted as one of the particles in this procedure it is natural that the jet definition should scale with the total hadronic energy $W^2$, but other choices are also possible. The proton remnant, which largely escapes in the beam pipe, can be represented by the momentum vector corresponding to the missing longitudinal momentum in the event.

A direct comparison of experimental jets with hard parton emissions in matrix element calculations can be made if the same resolution criterion is applied in the calculation. Thus, the theoretical cross-section is formulated in terms of the invariant masses of pairs of partons. The target remnant, with momentum $p_r = (1 - \xi)P$, is also included in order to automatically cut against collinear divergences in initial state emissions. The jet resolution $y_{cut}$ thereby defines the phase space for resolved parton emissions and sets the limits when integrating the matrix elements.

## 2.2 Cross-section formalism

Calculations of QCD matrix elements for jet production in DIS have been made since several years resulting in tree level diagrams up to order $\alpha_s^2$ (see [13] and references therein). For our purposes the complete $\mathcal{O}(\alpha_s^2)$ DIS 2+1 jet cross-section is needed, which have only recently been calculated (for one-photon exchange) by Brodkorb, Körner and Mirkes [13]. Prior to this result, the partial result obtained by Graudenz [14] by contracting the hadronic tensor with the metric tensor was available. The following results are mainly based on the complete calculation as implemented in the DISJET [15] program, but comparison is made with the earlier result as available in PROJET [16]. Since both calculations are performed in the $\overline{MS}$ scheme [17] our results for the scales also apply in that scheme.

Assuming pure photon exchange and integrating over the azimuthal angle $\phi$ between the parton plane and the lepton plane the cross-section can be written in the following general form,

$$\frac{d\sigma}{dxdy} = \frac{2\pi\alpha_{em}^2}{x^2 y^2 s} \left[ (1 + (1-y)^2)\sigma_{U+L} - y^2 \sigma_L \right] \qquad (2)$$

where $\sigma_{U+L}$ (or $F_2$) represents the helicity cross-section for unpolarised photons and $\sigma_L$ (or $F_L$) longitudinally polarised photons (the symbol $U$ stands for unpolarised transverse). The longitudinal structure function $F_L$ is a linear combination of the structure functions $F_1$ and $F_2$, $F_L = F_2 - 2xF_1$ which is zero in the Quark Parton Model (QPM) with spin 1/2 pointlike constituents (Callan-Gross relation). In terms of the hadronic tensor $H^{\mu\nu}$ we have

$$d\sigma_{U+L} = \left( -\frac{1}{2} g_{\mu\nu} + \frac{6x_p^2}{Q^2} p_\mu p_\nu \right) H^{\mu\nu}$$

$$d\sigma_L = \frac{4x_p^2}{Q^2} p_\mu p_\nu H^{\mu\nu}$$

where $p$ is the momentum of the incoming parton, $p = \xi P$ and $x_p \equiv Q^2/2pq = x/\xi$. Other linear combinations are also possible like the one used by Graudenz [14],

$$\frac{d\sigma}{dxdy} = \frac{2\pi\alpha_{em}^2}{x^2 y^2 s} \left[ \frac{1 + (1-y)^2}{2} \sigma_g + \frac{6(1-y) + y^2}{2} \sigma_0 \right] \qquad (3)$$

where $\sigma_g$ and $\sigma_0$ represents the 'metric' and 'non-metric' structure functions respectively,

$$d\sigma_g = -g_{\mu\nu} H^{\mu\nu}$$

$$d\sigma_0 = \frac{4x_p^2}{Q^2} p_\mu p_\nu H^{\mu\nu}$$

The two alternatives are related by $\sigma_g = 2\sigma_{U+L} - 3\sigma_L$ and $\sigma_0 = \sigma_L$.

## 3   Renormalisation scale choices

To make the dependence on the renormalisation scale $\mu_R$ explicit, the leading order (LO) and next-to-leading order (NLO) 2+1 jet cross section can be written

$$\frac{d\sigma_{2+1}^{LO}}{dx\,dQ^2} = A \frac{\alpha_s(\mu_R^2)}{2\pi} \qquad (4)$$

$$\frac{d\sigma_{2+1}^{NLO}}{dx\,dQ^2} = A \frac{\alpha_s(\mu_R^2)}{2\pi} \left[ 1 + \left( BN_f + C + b_o \ln(\mu_R^2/Q^2) \right) \frac{\alpha_s(\mu_R^2)}{2\pi} \right] \qquad (5)$$

where $A$, $B$ and $C$ are functions of $(x, Q^2, y_{cut}, \mu_F^2)$ after the dependences on the internal jet variables $(x_p, z, \phi)$ have been integrated out. The scale dependence of the strong coupling constant is given by a renormalisation group equation, the so called $\beta$-function, which to second order is given by

$$\frac{\partial}{\partial \ln \mu_R^2} \frac{\alpha_s(\mu_R^2)}{2\pi} = -b_0 \left( \frac{\alpha_s(\mu_R^2)}{2\pi} \right)^2 \left[ 1 + b_1 \frac{\alpha_s(\mu_R^2)}{2\pi} \right] \qquad (6)$$

where

$$b_0 = \frac{11}{6}N_c - \frac{1}{3}N_f$$

$$b_1 = \frac{\frac{17}{3}N_c^2 - \frac{5}{3}N_cN_f - C_FN_f}{2b_0}$$

and $N_c$ is the number of colours in QCD, $N_f$ is the number of light fermions and $C_F = (N_c^2 - 1)/2N_c$ is the Casimir factor. The commonly used standard solution to Eq. (6) is

$$\frac{\alpha_s(\mu_R^2)}{2\pi} = \frac{1}{b_0 \ln(\mu_R^2/\Lambda^2)} \left[1 - \frac{b_1 \ln\ln(\mu_R^2/\Lambda^2)}{b_o \ln(\mu_R^2/\Lambda^2)}\right] \quad (7)$$

where $\Lambda$ is the QCD scale parameter which sets the normalisation of the strong coupling. It is convenient to use the dimensionless rescaled renormalisation scale

$$\rho^2 = \frac{\mu_R^2}{Q^2} \quad (8)$$

such that a direct comparison is made to the naive choice of $Q^2$ as the scale in $\alpha_s$. As an example the renormalisation scale dependence of the LO and the NLO expressions for the cross-section are shown in Fig. 2. Although the result displayed is for specific values of $(x, Q^2, y_{cut})$, it illustrates a general behaviour. With changing $(x, Q^2, y_{cut})$ the curves are displaced, but their functional dependence of the renormalisation scale is unchanged. As an illustration the three different scale choices (calculated as discussed below) according to BLM, FAC and PMS are also shown in Fig. 2. The 'geometric' interpretation of the FAC and PMS scales can easily be seen whereas for the BLM scale there is no such simple interpretation.

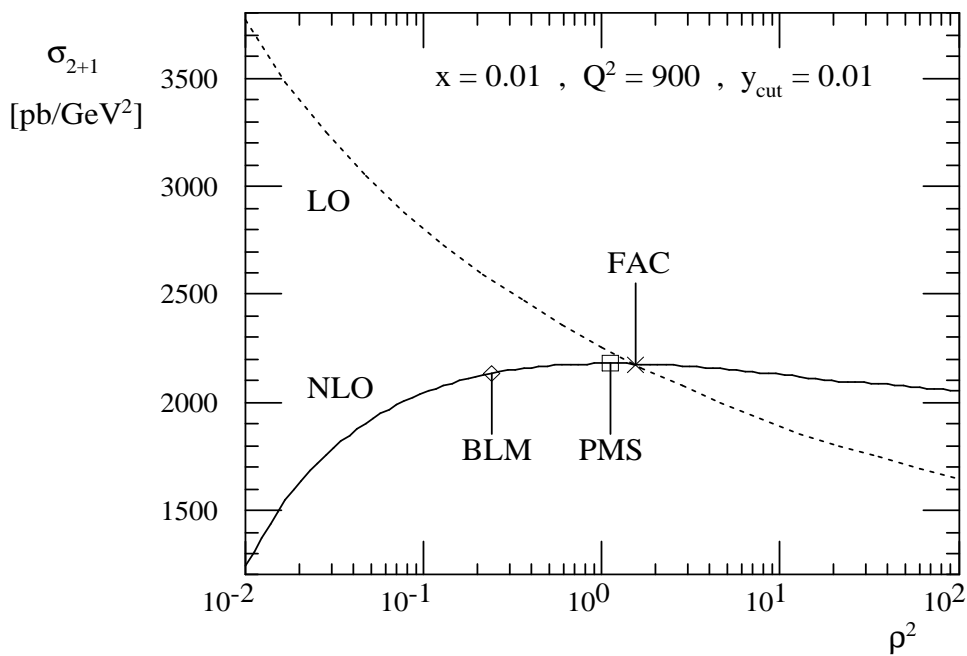

Figure 2: *The renormalisation scale ($\rho^2 = \mu_R^2/Q^2$) dependence of the 2+1 jet cross-section in leading order (dotted curve) and next-to-leading order (solid curve) at $x$, $Q^2$ and with $y_{cut}$ as indicated. The three different scale choices are also indicated.*

## 3.1 The BLM scale

The Brodsky-Lepage-Mackenzie (BLM) [8] method is inspired by QED where all fermion loops (vacuum polarisations) in the photon propagator are absorbed into the running of the coupling $\alpha_{em}$. The same condition is applied to QCD after the renormalisation scheme has been chosen, i.e. all the quark and gluon vacuum polarisations in the gluon propagator are absorbed into the running of $\alpha_s$. In NLO it is sufficient to keep track of the $N_f$ dependent terms since the vacuum polarisations have the form $b_0 K$, so the $N_c$-part follows automatically with the $N_f$-part. This means that all $N_f$ dependent terms showing up are a sign of bad bookkeeping and they should be absorbed into the running of $\alpha_s$. Although this principle is general it has to be applied with some caution in processes with gluon-gluon scattering in LO [8]. For the DIS 2+1 jet cross-section, there is however, no problem to find the renormalisation scale with the BLM method. The cross-section in Eq. (5) provides the condition

$$BN_f - \frac{1}{3}N_f \ln(\mu_R^2/Q^2) = 0 \tag{9}$$

from which the BLM scale is obtained as

$$\rho_{BLM}^2 = \exp(3B) \tag{10}$$

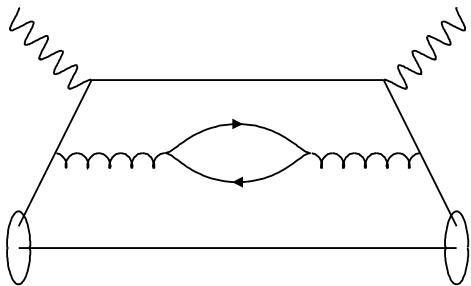

Figure 3: *The squared amplitude for the $N_f$ dependent term in the NLO 2+1 jet cross-section.*

To $\mathcal{O}(\alpha_s^2)$ the only process which gives an explicit $N_f$-dependence is illustrated in Fig. 3. The analytic form of this term can be obtained from the explicit cross-section formulas in [13, 14] and it is given by

$$B = \frac{\int dx_p\, dz\, d\phi\, A_{compton} \left[\frac{1}{3}\ln\left(\min\left\{\frac{1-x_p}{x_p}(1-z), \frac{1-x}{x}y_{cut}\right\}\right) - \frac{5}{9}\right]}{\int dx_p\, dz\, d\phi\, (A_{compton} + A_{fusion})} \tag{11}$$

Where $A_{compton}$ is the Born cross-section for the QCD Compton process and $A_{fusion}$ is the Born cross-section for the boson-gluon fusion process as seen in Fig. 4. (Comparing with Eq. (5) we have $A = \int dx_p\, dz\, d\phi\, (A_{compton} + A_{fusion})$.)

## 3.2 The FAC scale

A common way of judging the reliability of a series expansion in physics is to look at the relative size of the last term calculated. If this term is small then the expansion is considered as being trustworthy and if it is big then it is not. The method of Fastest Apparent

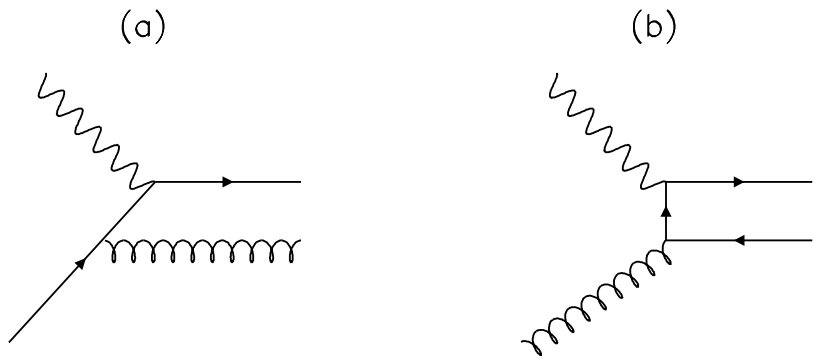

Figure 4: *Illustration of the two Born processes for 2+1 jet production on parton level: (a) QCD Compton or gluon radiation. (b) Boson-gluon fusion.*

Convergence (FAC) [9] makes use of this criterion by choosing the renormalisation scale so that the last term in the perturbative expansion equals zero. In our case this means that the FAC scale is obtained from the requirement that the $\alpha_s^2$ term in Eq. (5) is zero, i.e.

$$BN_f + C + b_o \ln(\mu_R^2/Q^2) = 0 \tag{12}$$

which gives

$$\rho_{FAC}^2 = \exp\left(-\frac{BN_f + C}{b_0}\right) \tag{13}$$

## 3.3 The PMS scale

The Principle of Minimal Sensitivity (PMS) [10] defines the renormalisation scale by demanding that $\partial R/\partial d\mu_R^2 = 0$ for any physical observable $R$. This choice is based on the knowledge that the complete perturbative series is renormalisation scale independent so one may try to impose the same condition on the truncated series. As showed by Stevenson in his original paper [10], demanding $\partial R/\partial \mu_R^2 = 0$ will also make the optimised NLO result for $R$ renormalisation scheme independent. In general the PMS principle should be applied both to the renormalisation scale and scheme. In that respect the PMS method differs from the BLM and FAC methods which only set the scale.

A convenient way of parameterising different renormalisation schemes is through the coefficients ($b_i$) in the $\beta$-function,

$$\frac{\partial \lambda}{\partial \ln \mu_R^2} = -b_0 \lambda^2 \left[1 + b_1 \lambda + b_2 \lambda^2 + ...\right] \tag{14}$$

where we have introduced the short hand notation

$$\lambda = \frac{\alpha_s(\mu_R^2)}{2\pi} \tag{15}$$

The first two coefficients, $b_0$ and $b_1$ are renormalisation scheme independent but the rest of them ($b_i, i > 1$) are not. The PMS principle then amounts to requiring

$$\frac{\partial R}{\partial \mu^2} = 0$$
$$\frac{\partial R}{\partial b_i} = 0 , \; i > 1$$

The last condition is only applied for the $b_i$ which are kept in the $\beta$-function (the rest of the $b_i$'s are set to zero). In our case this means that only the first condition has to be applied to the cross-section. Taking the derivative of Eq. (5) and using Eq. (6) gives

$$\frac{\partial \sigma_{2+1}^{NLO}}{\partial \ln \mu_R^2} = -Ab_0\lambda^3 \left[ b_1 + 2(1+b_1\lambda)\left(BN_f + C + b_o\ln(\mu_R^2/Q^2)\right)\right] \qquad (16)$$

The PMS scale is then obtained from the requirement

$$\frac{\partial \sigma_{2+1}^{NLO}}{\partial \ln \mu_R^2} = 0 \qquad (17)$$

which gives

$$b_1 + 2(1+b_1\lambda)\left(BN_f + C + b_o\ln(\mu_R^2/Q^2)\right) = 0 \qquad (18)$$

This is a transcendental equation so its solution can not be given as a closed expression but it can be solved numerically to arbitrary precision. For small $\lambda$, the $\mu_R$-dependence is mainly through the term $b_0 \ln(\mu_R^2/Q^2)$ in comparison to the $b_1\lambda$ term, and one can therefore obtain the following iterative solution,

$$\lambda^{(n+1)} = \frac{\alpha_s\left(Q^2 \exp\left[-\frac{BN_f+C}{b_0} - \frac{b_1}{2b_0(1+b_1\lambda^{(n)})}\right]\right)}{2\pi} \qquad (19)$$

with $\lambda^{(0)} = 0$. We have used the second iteration as the solution, i.e.

$$\rho_{PMS}^2 = \exp\left(-\frac{BN_f+C}{b_0} - \frac{b_1}{2b_0(1+b_1\lambda^{(1)})}\right) \qquad (20)$$

which is precise enough for our purposes. Numerically we have (for $N_f = 5$) $b_0 = \frac{23}{6}$ and $b_1 = \frac{58}{23}$, so for the small $\lambda \sim 0.02 - 0.03$ we are interested in the error is negligible. In passing we also note that comparing Eq. (13) and Eq. (20) we get the approximate relation $\rho_{PMS}^2 \simeq 0.72\,\rho_{FAC}^2$ (for $N_f = 5$).

## 3.4 Analytic expressions for the scales

It would be instructive to have useful analytic expressions for the scales, but this is prohibited mainly by the long and cumbersome expression for $C$. We have therefore resorted to numerical calculation of the parameters $A$, $B$ and $C$ to get the different scales.

There is also the question whether the scales should be calculated for the full 2+1 jet cross-section or the two Born processes separately. If one divides the cross-section into two parts, one for the $q\bar{q}$ final state and one for the $qg$, the following expressions for the BLM scales are obtained

$$\rho_{BLM,compton}^2 = \exp\left(\frac{\int dx_p\, dz\, d\phi\, A_{compton} \ln\left(\min\left\{\frac{1-x}{x}y_{cut}, \frac{1-x_p}{x_p}(1-z)\right\}\right)}{\int dx_p\, dz\, d\phi\, A_{compton}} - \frac{5}{3}\right) \qquad (21)$$

$$\rho_{BLM,fusion}^2 = 1$$

In the region of phase space where $\frac{1-x_p}{x_p}(1-z)$ does not contribute we get the interpretation that the renormalisation scale is simply given by the cut-off in the jet definition, $y_{cut}W^2$, times a numerical factor. Theoretically one can have different scales for different processes. In fact, there is nothing that says that one should even have the same scale in the $\alpha_s$ and $\alpha_s^2$ contributions. In principle one could therefore also investigate the scale dependence of these two Born processes separately. We have chosen not to pursue this line mainly because experimentally one is first of all interested in the inclusive 2+1 jet cross-section, such that the extra uncertainties of quark and gluon jet identification does not enter. Furthermore, the use of different scales for the two Born processes will make it difficult to find a common $y_{cut}$ where the theoretical uncertainty is small. Having different $y_{cut}$ for the two Born processes would be very awkward experimentally, since it would require a separation of gluon jets from quark jets on an event-by-event basis before the jet reconstruction with resolution $y_{cut}$ has been made!

## 3.5 The factorisation scale

The dependence on the factorisation scale $\mu_F^2$ comes both from the parton densities and explicit $\alpha_s^2$ terms of the form $D \ln \mu_F^2/Q^2$. Thus, the $\mathcal{O}(\alpha_s^2)$ DIS 2+1 jet cross-section takes the general form

$$\frac{d\sigma_{2+1}^{NLO}}{dx\, dQ^2} = A\,\lambda \left[1 + \left(B(\mu_F^2)N_f + C(\mu_F^2) + b_o \ln(\frac{\mu_R^2}{Q^2}) + \left(D(\mu_F^2)b_0 + E(\mu_F^2)\right)\ln(\frac{\mu_F^2}{Q^2})\right)\lambda\right] \tag{22}$$

where

$$D(\mu_F^2) = \frac{\int dx_p\, dz\, d\phi\, A_{fusion}(\mu_F^2)}{\int dx_p\, dz\, d\phi\, (A_{compton}(\mu_F^2) + A_{fusion}(\mu_F^2))} \tag{23}$$

The factorisation scale dependence of the parton density functions is given by the GLAP [18] evolution equations

$$\frac{\partial f_q(\xi,\mu_F^2)}{\partial \ln \mu_F^2} = \frac{\alpha_s(\mu_F^2)}{2\pi} \int_\xi^1 \frac{dz}{z}\left[P_{qq}\left(\frac{\xi}{z}\right)f_q(z,\mu_F^2) + P_{qG}\left(\frac{\xi}{z}\right)f_G(z,\mu_F^2)\right]$$

$$\frac{\partial f_G(\xi,\mu_F^2)}{\partial \ln \mu_F^2} = \frac{\alpha_s(\mu_F^2)}{2\pi} \int_\xi^1 \frac{dz}{z}\left[P_{Gq}\left(\frac{\xi}{z}\right)f_q(z,\mu_F^2) + P_{GG}\left(\frac{\xi}{z}\right)f_G(z,\mu_F^2)\right]$$

where $P_{qq}$ etc are the splitting functions which are known to NLO [19]. With this information it should be possible to minimise the factorisation scale dependence of the cross-section and therefore extract the PMS scale. However, for the FAC and BLM scales one needs some kind of analytic form for the structure functions so that the $\alpha_s^2$- and $N_f$-contributions can be identified. In this first step towards a complete understanding of the theoretical uncertainties in the DIS 2+1 jet cross-section we have limited ourselves to the renormalisation scale dependence. Only some preliminary numerical exercises will be made towards a study of the factorisation scheme and scale dependences that is needed to make the picture complete.

# 4  Numerical results

In order to calculate the renormalisation scales and the corresponding 2+1 jet cross-section the quantities $A$, $B$ and $C$ in Eq. (5) have been calculated numerically. This has been made for fixed $(x, Q^2, y_{cut})$ using the Monte Carlo program DISJET 1.0 [15], which is based on the complete $\mathcal{O}(\alpha_s^2)$ calculation [13]. For comparison we have also used the Monte Carlo program PROJET 3.6 [16] which only includes the $\alpha_s^2$ term proportional to $\sigma_g$ in Eq. (3). The calculations were made for the HERA design energy, 30 $GeV$ electrons on 820 $GeV$ protons giving a cms energy of $\sqrt{s} = 314\ GeV$. The dependence on the main DIS variables $x, Q^2$ and the jet resolution cut-off $y_{cut}$ will be of primary importance. We have therefore chosen $(x, Q^2)$-points to cover most of the kinematic range accessible at HERA as can be seen in Fig. 5. The $x$-values considered are 0.0001, 0.0003, 0.001, 0.003, 0.01, 0.03, 0.1, 0.3 and 0.9 and the $Q^2$-values are 9, 90, 900 and 9000 $GeV^2$.

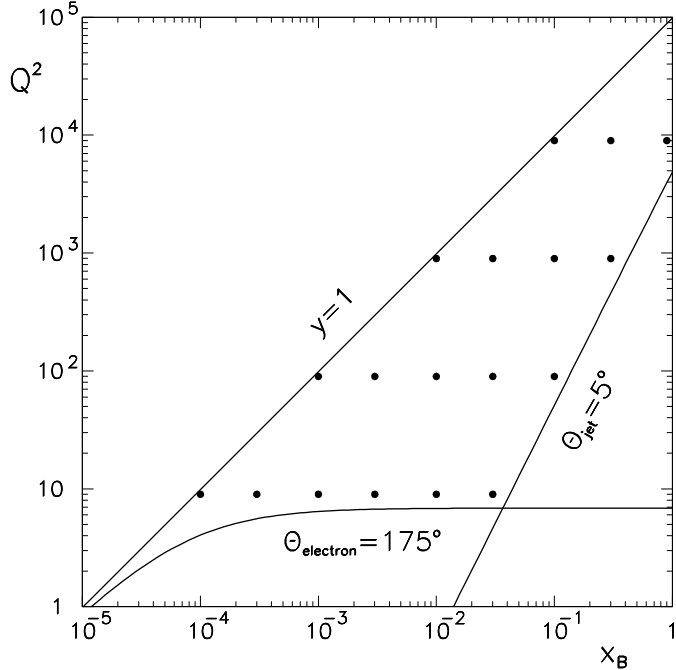

Figure 5: $(x, Q^2)$-points in the HERA kinematic range, with polar angle acceptance limits for the scattered electron and quark-jet (in QPM).

The $y_{cut}$ values have been chosen to cover what can be considered as a possible region, i.e. $y_{cut} = 0.005, 0.01, 0.02, 0.04, 0.08$. For $y_{cut}$ smaller than $\sim 0.01$ one expects the perturbative series to break down due to large logarithms of $y_{cut}$ causing $\alpha_s \ln y_{cut} \ll 1$ to be invalid. This problem can in principle be solved by a resummation of such terms (similar to the leading $\ln Q^2$ resummation in the GLAP equations.) For $y_{cut}$ larger than $\sim 0.08$ [15] the neglect of terms proportional to $y_{cut}$ in the cross-section calculations is no longer expected to be valid.

For the required proton structure functions we have used parton density parameterisations from the package PAKPDF 2.1 [20], where we select the latest parameterisation by the CTEQ collaboration [21], set CTEQ2M which is a NLO fit including the first HERA data. The structure functions are given in the $\overline{MS}$ scheme and for the QCD scale parameter we have used their fitted value $\Lambda_{\overline{MS}}^{(4)} = 0.213\ GeV$. The choice of parton density functions should not play a big role for the resulting scales and cross-sections since the jet definition introduces a cut-off against small momentum fractions $\xi$,

$$\xi \geq x + y_{cut}(1 - x) \qquad (24)$$

and all modern parameterisations more or less agree for $\xi \geq 0.01$. The factorisation scale has been set to $Q^2$ which is arbitrary, but convenient for our purposes. Some estimates with other choices will be given below.

## 4.1 Resulting renormalisation scales

A selection of the scales obtained using the complete second order 2+1 jet cross-section (DISJET) is shown in Fig. 6. For a complete collection of all the scales calculated see Table 3 in appendix A.

Fig. 6 demonstrates that the scales obtained with the different methods may differ quite substantially. The FAC and PMS scales are quite close, as expected from their definition (and discussed at the end of section 3.3). The BLM scale has, however, a different behaviour and in some $(x, Q^2, y_{cut})$ points it differs by orders of magnitude. The following trends are clear from studying Fig. 6.

- The BLM scale decreases with increasing $x$, whereas the FAC and PMS scale do not have any simple $x$ dependence.

- All three methods gives $\mu_R^2$ approximately proportional to $Q^2$.

- The BLM scale increases with increasing $y_{cut}$, whereas the FAC and PMS scales increase with increasing $y_{cut}$ for small $x$ but increase with decreasing $y_{cut}$ for large $x$.

Comparing the scales one sees that they cross around $y_{cut} = 0.01$ for small $x$ and around $y_{cut} = 0.04$ for large $x$. The increase of the FAC and PMS scales observed for large $x$ at small $y_{cut}$ is unphysical and indicates that the perturbative expansion is breaking down. On the other hand, the BLM scale behaves in a way which is what we expect physically, it increases when $y_{cut}$ is increased and decreases when $x$ is increased. This is a natural consequence of the jet-definition which is a measure of the typical virtualities in the process. We are therefore led to believe more in the BLM scale than the FAC and PMS scales even though the perturbative expansion can be questioned when the PMS and FAC scales start to grow in this way.

The increase of the FAC and PMS scales at small $x$ and large $y_{cut}$ is different from what has been obtained by Kramer and Lampe [22] for 3-jet rates in $e^+e^-$ at LEP energies. Their result seems to be more like what is obtained for large $x \sim \frac{1}{3}$ as seen in Fig. 7. At first one might expect that the results in DIS should be similar to $e^+e^-$ since the processes are related on parton level through crossing. There are however some important differences. The $e^+e^-$ calculation is obtained by contracting the hadronic tensor with the metric one. This means that the 'non-metric' part defined as in Eq. (3) is not present in the $e^+e^-$ calculation. Furthermore, at small $x$ the boson-gluon fusion process starts to dominate in DIS due to the increase of the gluon density. It is presumably the large contribution of this process to the 'non-metric' longitudinal part which is the main difference between DIS and $e^+e^-$ in this context.

The scales have also been calculated using PROJET [16] which only includes the contraction of the hadronic tensor with the metric tensor for the $\mathcal{O}(\alpha_s^2)$ cross-section. Therefore only the part proportional to $\sigma_g$ in Eq. (3) can be used to determine the scales.

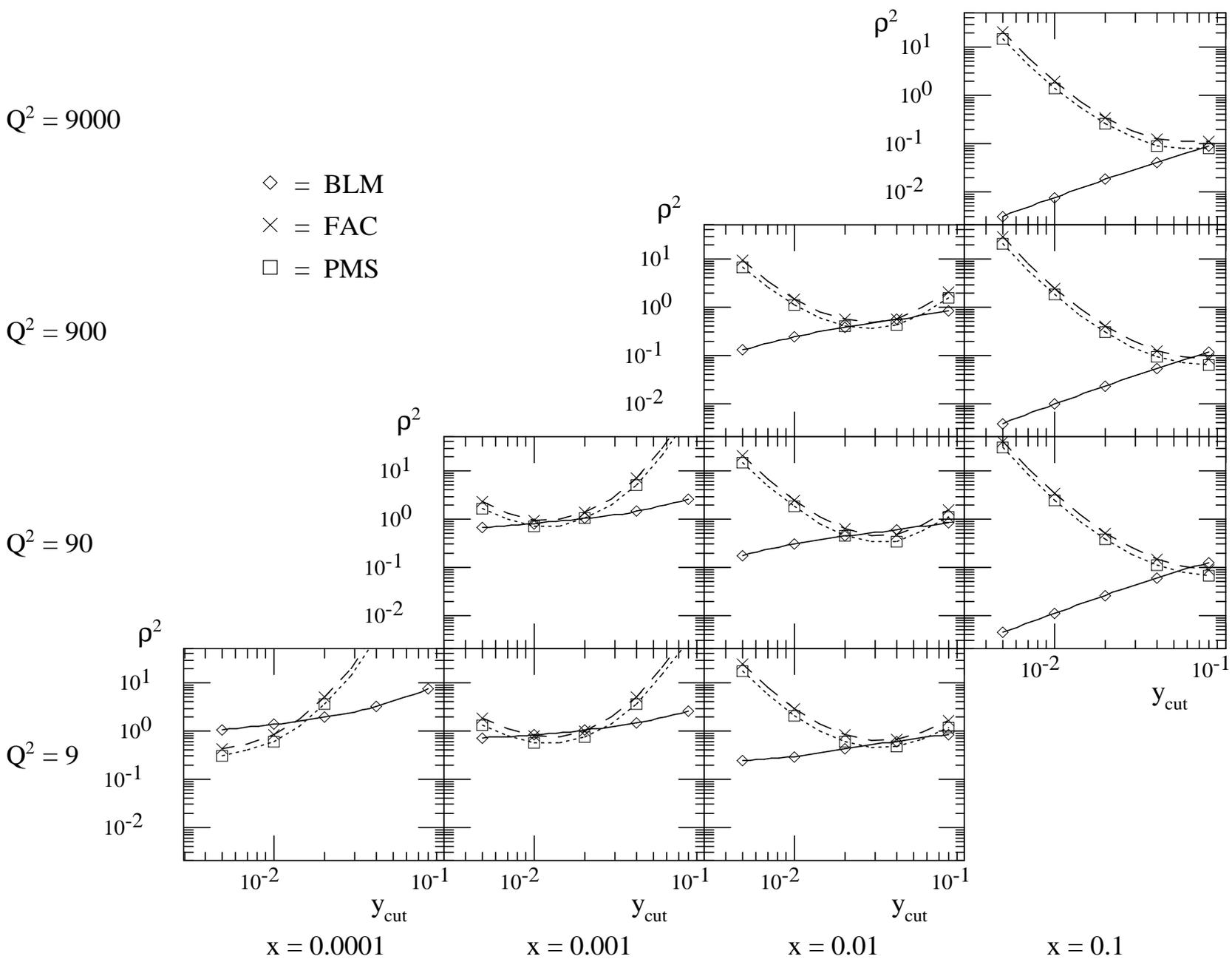

Figure 6: *The renormalisation scales obtained according to the BLM (solid), FAC (dashed) and PMS (dotted) methods. The lines connect the points actually calculated.*

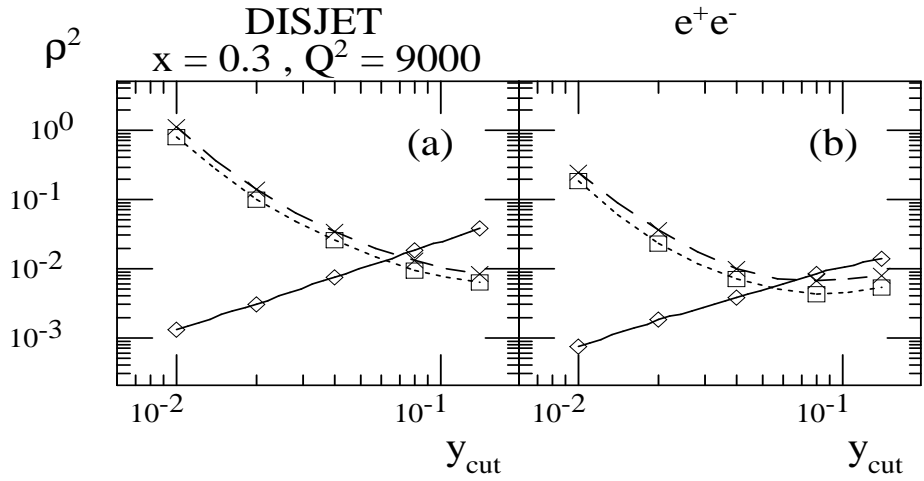

Figure 7: *Comparison of the scales obtained from the BLM (solid), FAC (dashed) and PMS (dotted) methods for (a) 2+1 jet production in DIS at $x = 0.3$, $Q^2 = 9000~GeV^2$ and (b) 3-jet production in $e^+e^-$ as given in ref. [22].*

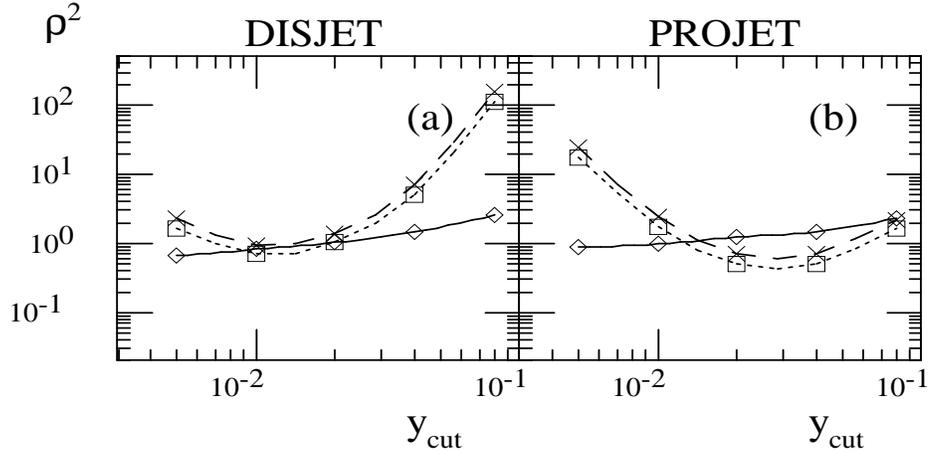

Figure 8: *Comparison of the scales obtained in $x = 0.001$, $Q^2 = 90 GeV^2$ according to the BLM (solid), FAC (dashed) and PMS (dotted) methods with (a) the complete $\alpha_s^2$ cross-section (DISJET) and (b) the partial result (PROJET).*

The values for the BLM scale are almost the same as those obtained with DISJET but the FAC and PMS scales are quite different. As an example the scales obtained for $x = 0.001$ and $Q^2 = 90 GeV^2$ are shown in Fig. 8. The differences observed can be understood from the fact that the $\alpha_s^2$ corrections to the 'non-metric' part ($\sigma_0$) of the cross-section are relatively larger than the $\alpha_s^2$ corrections to $\sigma_g$. The differences at large $x$ are much smaller than at small $x$ which is also natural since the 'non-metric' longitudinal part of the cross-section is relatively larger at small $x$ and therefore gives larger impact there. For the BLM scale, on the other hand, the $N_f$-dependent terms are the same for the unpolarised and longitudinally polarised contributions, so that the resulting scale is the same. Comparing the scales obtained with PROJET with the ones from $e^+e^-$ one sees in this case a closer agreement for all $x$, as expected when the 'non-metric' part of the cross-section is neglected.

## 4.2 Cross-sections with different scale choices

With the scales calculated it is easy to get the cross-sections from Eq. (5) for the respective scales. The results are shown in Fig. 9 for the same $(x, Q^2)$-points as for the scales in Fig. 6. The cross-section is seen to increase with increasing $x$ and decreasing $y_{cut}$ as a consequence of an increased phase space for resolved jets as given by the jet resolution ($s_{ij} \geq y_{cut}\frac{1-x}{x}Q^2$). The strong decrease of the cross-section with $Q^2$ is mainly the trivial propagator effect from the photon exchange.

In general we observe smaller differences between the cross-sections than between the scales and the decrease is more than just due to the logarithmic scale dependence of the running coupling constant. This is also what we expect from Fig. 2 where the scale dependence of the LO and NLO cross-sections can be compared. However, for large $x$ and small $y_{cut}$ the cross-section for the BLM scale diverges as the $\alpha_s^2$ corrections become negative and larger in magnitude than the Born term.

## 4.3 Uncertainty due to renormalisation scale

It is important to estimate the error in the theoretical calculation of the cross-section in order to use it for precision measurements of $\alpha_s$ etc. There is, however, no 'correct' way of doing this and one is therefore left with some arbitrariness or prejudice (unless the exact answer is known). In our case we only know the first two terms in a series that may even not converge properly (see however the fourth paper listed in [10] for a discussion on the possible convergence of 'optimised' approximations and the Borel summability of QCD and also [23] for a comparison of the NLO and NNLO optimised scales in $R_{e^+e^-}$).

As already stated, the FAC method is based on one of the more common methods for estimating errors, the *apparent convergence* of the series. If this criterion is used then the error would be zero for the FAC scale, $\frac{b_1 \lambda_{PMS}}{2(1+b_1\lambda_{PMS})}$ for the PMS scale and $(C + \frac{33}{2}B)\lambda_{BLM}$ for the BLM scale. One could also be very conservative and compare the three LO and NLO cross-sections simultaneously and assign the global difference as an error. This is, however, unnecessarily pessimistic since it ignores some knowledge.

Another possibility is to look at the logarithmic derivative of the cross-section with respect to the renormalisation scale, $\partial \ln \sigma / \partial \ln \mu_R^2$. By definition the error would then be zero for the PMS scale, $b_1 \lambda_{FAC}^2$ for the FAC scale and

$$\left.\frac{\partial \ln \sigma}{\partial \ln \mu_R^2}\right|_{BLM} = \frac{\lambda_{BLM}^2 \left[b_1 + 2(1 + b_1 \lambda_{BLM})\left(C + \frac{33}{2}B\right)\right]}{\left[1 + \left(C + \frac{33}{2}B\right)\lambda_{BLM}\right]} \quad (25)$$

for the BLM scale. If the renormalisation scale dependence is small, then the situation is at least more favourable than if it is large. In the latter case it is clear that the theoretical uncertainty is high, whereas in the former one it can be small even though this cannot be said with any certainty. In other words, it is a necessary but not sufficient condition that the renormalisation scale dependence is small.

It has been proposed by Brodsky and Lu [24] that $\partial \ln \sigma / \partial \ln \mu_R^2|_{BLM}$ (*i.e.* evaluated at $\mu_R^2 = \rho_{BLM}^2 Q^2$) is a suitable measure of the theoretical uncertainty. The argumentation for this is as follows. First of all the BLM scale should be chosen since this is the only scale which behaves in a physical way. Secondly $\partial \ln \sigma / \partial \ln \mu_R^2$ is a measure of the renormalisation scale dependence which indicates the size of the uncalculated terms. Therefore this has to be small and the ideal case would thus be when the BLM and PMS scales coincide.

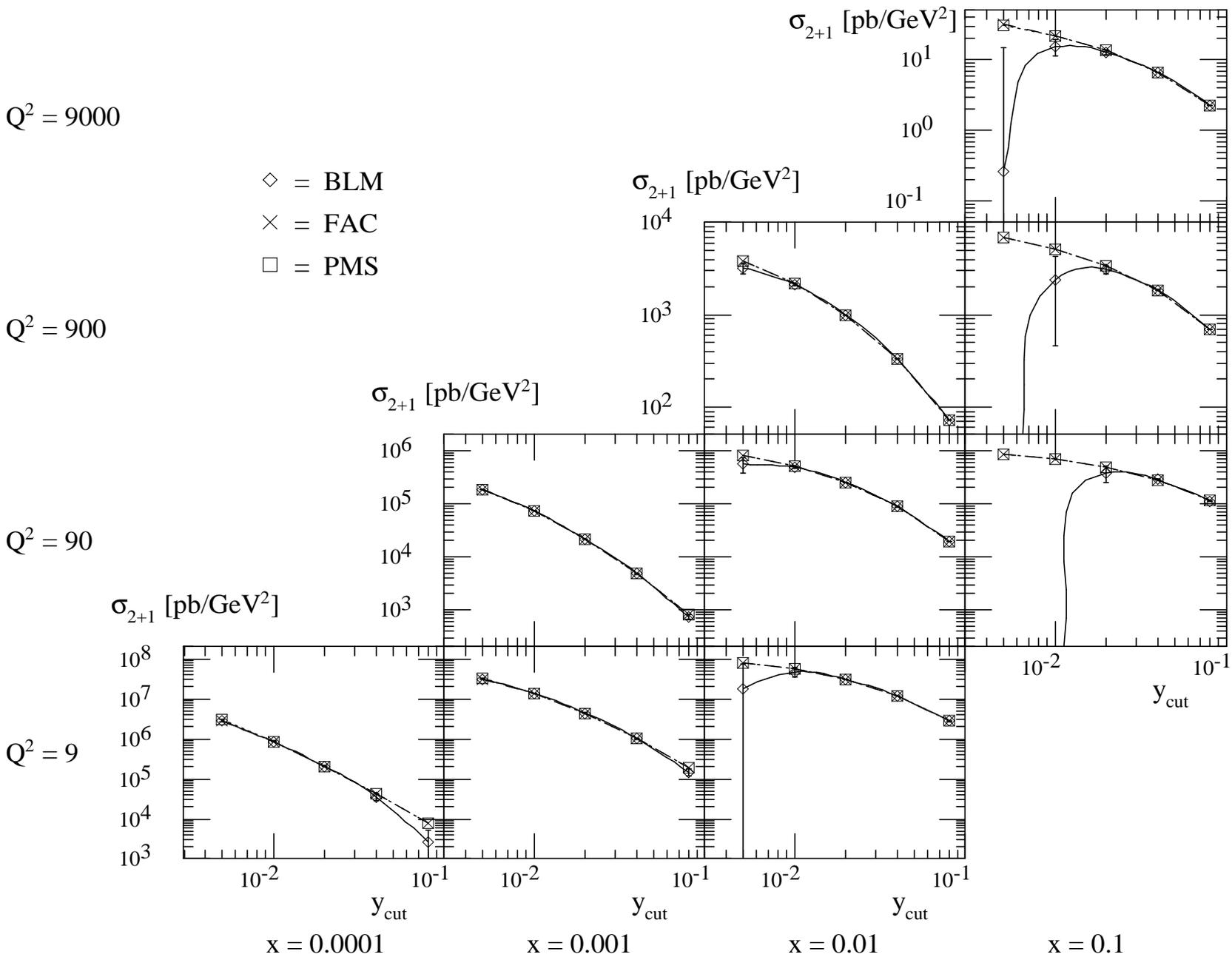

Figure 9: *DIS 2+1 jet cross-sections as obtained for the different scales. The error bars for the BLM scale corresponds to the derivative $\partial \ln \sigma / \partial \ln \mu_R^2$ in this point*

In Fig. 9 this error estimate is plotted for the points referring to the BLM scale and it is also given in Table 1. Both the figure and the table shows that the theoretical uncertainty becomes larger when $Q^2$ is decreased and $x$ and $y_{cut}$ are held fixed. This cannot be concluded from just comparing the scales in Fig. 6. The explanation is mainly that the logarithmic derivative (Eq. (25)) is proportional to $\alpha_s^2$ so that the running of the coupling gives this effect.

$$\partial \ln \sigma / \partial \ln \mu_R^2 |_{BLM}$$

| $x$ | $Q^2$ $[GeV^2]$ | $y_{cut}$ | | | | |
|---|---|---|---|---|---|---|
| | | .005 | .01 | .02 | .04 | .08 |
| 0.0001 | 9 | **0.050** | **0.032** | **0.027** | 0.185 | 0.917 |
| 0.0003 | 9 | **0.027** | **0.033** | **0.004** | 0.110 | 0.440 |
| 0.001 | 9 | **0.038** | **0.018** | **0.016** | **0.046** | 0.215 |
| 0.001 | 90 | **0.024** | **0.003** | **0.001** | **0.029** | 0.103 |
| 0.003 | 9 | 0.238 | **0.029** | **0.020** | **0.006** | 0.101 |
| 0.003 | 90 | 0.085 | **0.014** | **0.009** | **0.002** | **0.048** |
| 0.01 | 9 | 3.027 | 0.256 | **0.021** | **0.013** | **0.018** |
| 0.01 | 90 | 0.334 | 0.065 | **0.001** | **0.013** | **0.006** |
| 0.01 | 900 | 0.132 | **0.032** | **0.001** | **0.004** | **0.009** |
| 0.03 | 9 | – | – | 0.227 | **0.003** | **0.028** |
| 0.03 | 90 | 4.257 | 0.282 | **0.046** | **0.013** | **0.013** |
| 0.03 | 900 | 0.605 | 0.114 | **0.020** | **0.007** | **0.009** |
| 0.1 | 90 | – | – | 0.366 | **0.032** | **0.022** |
| 0.1 | 900 | – | 0.808 | 0.114 | **0.014** | **0.012** |
| 0.1 | 9000 | 55.21 | 0.271 | 0.068 | **0.014** | **0.001** |
| 0.3 | 900 | – | – | 0.621 | 0.069 | **0.019** |
| 0.3 | 9000 | – | 1.693 | 0.170 | **0.032** | **0.012** |
| 0.9 | 9000 | – | 1.442 | 0.134 | 0.174 | 0.164 |

Table 1: *Renormalisation scale uncertainty as defined by $\partial \ln \sigma / \partial \ln \mu_R^2 |_{BLM}$, with bold face to show the region with better than 5% theoretical precision. (Missing values indicate negative cross-section, cf. Fig. 9.)*

## 4.4 Preferred $y_{cut}$-values

With the theoretical uncertainty defined in the above described way one can exploit its dependence on the jet resolution ($y_{cut}$). It is easy to see from Table 1 that by choosing $y_{cut}$ in a suitable way one can always make the theoretical uncertainty small. Letting, e.g., $\partial \ln \sigma / \partial \ln \mu_R^2 |_{BLM} < 5\%$ define a small theoretical error gives the acceptable $y_{cut}$ values indicated with bold face in Table 1, *i.e.* for small $x$ one should use $y_{cut} \sim 0.01$ and for large $x$ one should use $y_{cut} \sim 0.04$.

One can also see from Table 1 that there is an interval of useful $y_{cut}$ values, especially at small $x$ there is a clear upper limit which is a new feature previously not encountered.

The lower limit can be understood from the need to resum large logarithms of $y_{cut}$ and the upper limit is probably due to the neglect of terms proportional to $y_{cut}$ in the calculations which makes the $\alpha_s^2$ corrections too much negative.

We note that this dependence on $y_{cut}$ is actually beneficial from an experimental point of view. First of all, a smaller $y_{cut}$ gives a larger rate and thereby statistics. Secondly, to be able to extract the gluon density at small $\xi$ [5] one wants to use a $y_{cut}$ as small as possible in accordance with Eq. (24).

## 4.5 BLM scale: parameterisation and analytic expression

Since the BLM scale has been shown to be almost strictly proportional to $Q^2$ it is natural to perform a fit in $x$ with all $Q^2$-points taken into account. The fit has been made for $y_{cut} = 0.02$ since this seems to be the best compromise if one insists on using the same $y_{cut}$ in the whole kinematic range. The following function has been fitted

$$\rho^2 = \exp(A + B \ln x + C \ln^2 x + D \ln^3 x) \tag{26}$$

with the result $A = -9.25$, $B = -3.37$, $C = -0.425$, $D = -0.0191$. The fit can bee seen

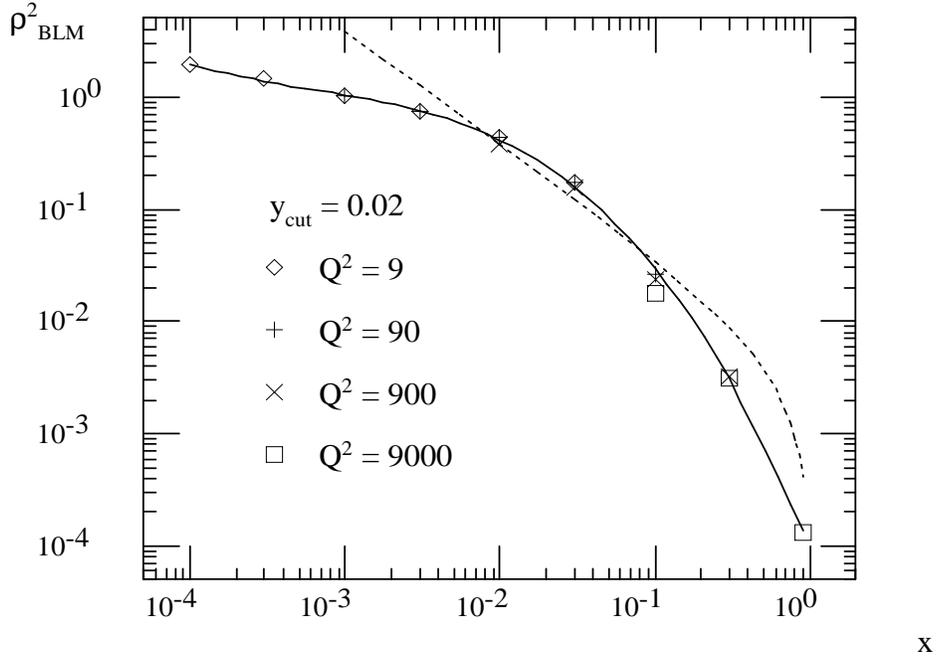

Figure 10: *Fit (solid) of $\rho^2_{BLM}$ using Eq. (26) for $y_{cut} = 0.02$ using all $x, Q^2$-points together with the analytic expression Eq. (27) (dotted).*

in Fig. 10 together with the analytic expression

$$\rho^2 = \exp\left(-\frac{5}{3}\right) y_{cut} \frac{1-x}{x} \tag{27}$$

The latter is expected to hold when the Compton process dominates and the term $\frac{1-x_p}{x_p}(1-z)$ does not contribute in the integration in Eq. (21). The deviations from these conditions explain the behaviour of our numerically calculated BLM scale. At small $x$ the fusion diagram dominates and therefore we expect the BLM scale to be close to $Q^2$. At large $x$, the expression $\frac{1-x_p}{x_p}(1-z)$ is smaller than $\frac{1-x}{x} y_{cut}$ in parts of the phase space and consequently the BLM scale gets smaller due to minimum condition in Eq. (21).

## 4.6 Comparison of $Q^2$ and the BLM scale

To see the potential effects of using the BLM scale instead of $Q^2$ as renormalisation scale, the fit to the BLM scale given above has been used to calculate the 2+1 jet cross-section in two $(x, Q^2)$-areas for $y_{cut} = 0.02$. The calculation has been made with the DISJET program for a high-$Q^2$ and a low-$Q^2$ bin with the following limits,

$$\begin{aligned} \text{low-}Q^2: \quad .0001 &< x < .02 \\ .05 &< y < .95 \\ 10 &< Q^2 < 100 \end{aligned}$$

$$\begin{aligned} \text{high-}Q^2: \quad .001 &< x < .1 \\ .05 &< y < .95 \\ 100 &< Q^2 < 9400 \end{aligned}$$

The resulting cross-sections are given in Table 2 and show very small differences ($\sim 1\%$) between the two scales.

$\sigma_{2+1}\ [pb]$

|  | low $Q^2$ | high $Q^2$ |
|---|---|---|
| $\mu_R^2 = Q^2$ | 1269 | 472 |
| $\mu_R^2 = \rho_{BLM}^2 Q^2$ | 1257 | 479 |

Table 2: *Cross-sections, in pb, for the low and high $Q^2$ bins.*

However, it is important to keep in mind that even if $\sigma(\mu_R^2 = Q^2) \sim \sigma(\mu_R^2 = \rho_{BLM}^2 Q^2)$, the $\Lambda_{QCD}$ extracted from analysing the running of the coupling will be different in the two cases. As an illustration, consider a measured 2+1-jet cross-section in one $(x, Q^2)$-point. With the quantities $A$, $B$ and $C$ calculated it is then possible to extract $\alpha_s$ once the renormalisation scale is chosen. To a first approximation the extracted value of $\alpha_s$ is independent of $\mu_R^2$ since the sum $BN_f + C + b_0 \ln(\mu_R^2/Q^2)$ is only weakly dependent on $\mu_R^2$. With $\alpha_s$ given, the value of $\Lambda_{QCD}$ will therefore scale with $\rho$ since $\alpha_s \sim 1/\ln(\mu_R^2/\Lambda_{QCD}^2)$. To be more quantitative and taking the $\mu_R^2$-dependence into account, assume that

$$\frac{d\sigma_{2+1}^{NLO}}{dx\, dQ^2}(x = 0.1,\, Q^2 = 900,\, y_{cut} = 0.04) = 1782\, pb/GeV^2$$

has been measured. Depending on whether $Q^2$ or $\rho_{BLM}^2 Q^2$ is used as scale one gets the following results

$$\begin{aligned} \mu_R^2 = Q^2: \quad \alpha_s = 0.1307 &\Rightarrow \Lambda_{\overline{MS}}^{(4)} = 213\, MeV \\ \mu_R^2 = \rho_{BLM}^2 Q^2: \quad \alpha_s = 0.1370 &\Rightarrow \Lambda_{\overline{MS}}^{(4)} = 74\, MeV \end{aligned}$$

where $\rho_{BLM}^2 = 0.05432$ has been used. One sees that taking the $\mu_R^2$-dependence from $b_0 \ln(\mu_R^2/Q^2)$ into account makes $\Lambda_{BLM} > \rho_{BLM}\Lambda_{Q^2} = 50\, MeV$.

## 4.7 Factorisation scale effects

Above we have only considered the case $\mu_F^2 = Q^2$. A complete analysis of the scale ambiguities in the DIS 2+1 jet cross-section would also have to include the effects of the factorisation scale uncertainty. Referring to Eq. (22) we see that $\mu_F^2 \neq Q^2$ will introduce extra terms, both $N_f$-dependent and $N_f$-independent, in addition to the $\mu_F$-dependence in the parton densities. A change of the factorisation scale will influence the renormalisation scale to an extent that depends on the factorisation scheme through the absorption into parton densities. As a rough estimate of the influence of the factorisation scale on our renormalisation scale results we have simply redone the determination of the latter by using Eq. (22) with $\mu_F^2 = 0.1\,Q^2$ and $\mu_F^2 = 10\,Q^2$.

In short the effects can be described as follows. First of all the general $(x, Q^2, y_{cut})$-dependencies are not affected, but the renormalisation scales are shifted when the factorisation scale is changed. For large $x$ the effects are small but for small $x$ they are quite considerable (factor 4–5). This big effect is more of a worst case than a realistic estimate. Physically we expect the factorisation scale to be of the same order as the renormalisation scale and since $\mu_R^2 \sim Q^2$ we do not expect the factorisation scale to be very different. As indicated, part of the $\mu_F$-dependence should presumably be absorbed against other terms from the structure functions and thus not contribute or affect the renormalisation scale. It is clear, however, that this deserves a separate study to obtain a complete estimate of the theoretical uncertainty.

## 5 Summary and conclusions

In this paper we have investigated the renormalisation scale dependence of the DIS 2+1 jet cross-section and three different methods for how to fix the renormalisation scale, the BLM, FAC and PMS methods. The different scales are obtained in the following ways: the BLM scale by absorbing all $N_f$-dependent terms into the running of $\alpha_s$, the FAC scale by requiring that the LO and NLO cross-sections coincide and the PMS scale by minimising the renormalisation scale dependence of the cross-section.

The scales were calculated using the complete $\mathcal{O}(\alpha_s^2)$ cross-section with the jets defined according to the JADE algorithm, $s_{ij} \geq y_{cut}W^2$, for representative $x$, $Q^2$ and $y_{cut}$ values in the HERA kinematic range. The scales are approximately proportional to $Q^2$ whereas the $x$ and $y_{cut}$ dependencies are more complicated. The BLM scale increases with decreasing $x$ and increasing $y_{cut}$ which is what we expect physically since the typical invariant mass of pairs of partons goes as $y_{cut}W^2 = y_{cut}Q^2\frac{1-x}{x}$ from the jet-definition. The FAC and PMS scales also increase with decreasing $x$ for large $y_{cut}$, but for small $y_{cut}$ the trend is the opposite. There is also an unphysical increase of the FAC and PMS scales for large $x$ when $y_{cut}$ is decreased. This indicates that the BLM scale is a better choice from a physical point of view.

The scales obtained using the complete $\mathcal{O}(\alpha_s^2)$ cross-section are also compared with the ones obtained using the partial cross-section from the contraction with the metric tensor and earlier results on 3-jet production in $e^+e^-$ [22]. For large $x$ the scales more or less agree and behave in a similar way as a function of $y_{cut}$, but for small $x$ the FAC and PMS scales from the complete DIS 2+1 jet cross-section behaves in a different way. The difference is mainly due to the dominance of boson-gluon fusion and, in particular, its contribution to the 'non-metric' longitudinal part of the cross-section which is only present in the complete $\mathcal{O}(\alpha_s^2)$ DIS cross-section.

To estimate the theoretical uncertainty due to the renormalisation scale dependence we use the logarithmic derivative of the cross-section with respect to the renormalisation scale evaluated for the BLM scale. By choosing an appropriate value of $y_{cut}$ it is possible to make the uncertainty defined in this way small. The range of useful $y_{cut}$ values is $y_{cut} \sim 0.01 - 0.04$ with the smaller value for small $x$ and the larger one for large $x$. Thus, a reliable theoretical result sets not only a lower limit on $y_{cut}$, but also an upper one.

Our results on useful $y_{cut}$ values can be translated into other jet-definitions which also use invariant mass cut-offs, e.g. $s_{ij} \geq y'_{cut} M^2$. The suitable values of $y'_{cut}$ are then simply given by rescaling of our results, $y'_{cut} = \frac{W^2}{M^2} y_{cut}$.

Choosing $y_{cut}$ values as indicated above the differences between the cross-sections for the different scales is very small, $\sim 1\%$. This can be compared with the difference between the LO and NLO cross-sections using $Q^2$ as scale which is in the order of $\sim 10\%$. It is also interesting to note that for these $y_{cut}$-values the preferred scales are in the order of $\mu_R^2 \sim Q^2$ for small $x$ and $\mu_R^2 \sim 0.1\,Q^2$ for large $x$.

A fit of the BLM scale ($\rho_{BLM}^2$) as a function of $x$ has been made for $y_{cut} = 0.02$ using all $Q^2$ points. This fit is then used to calculate the cross-section for a low and a high $Q^2$ region in the $(x, Q^2)$-plane. The difference in the cross-section obtained using the BLM scale and the simple $Q^2$ as scale is small, $\sim 1\%$. Nevertheless it is important to use the correct scale if the running of $\alpha_s$ is to be analysed or a proper value of $\Lambda_{QCD}$ extracted.

A complete analysis of the scale ambiguities in DIS would also have to include the factorisation scale dependence. The effects of using $\mu_F^2 \neq Q^2$ have been estimated crudely by calculating the renormalisation scales for two different choices, $\mu_F^2 = 0.1\,Q^2$ and $\mu_F^2 = 10\,Q^2$. The conclusions on suitable $y_{cut}$-values and cross-section differences are the same, but the numerical values of the scales are changed.

In conclusion, we have demonstrated that with an appropriate jet definition it is possible to control the renormalisation scale uncertainty of the DIS 2+1-jet cross-section such that precision tests of QCD are facilitated.

**Acknowledgements:** We are grateful to S. Brodsky for several enlightening and helpful discussions and to Yu. Dokshitzer and G. Kramer for comments on the manuscript. We would also like to thank E. Mirkes for helpful communications regarding the DISJET Monte Carlo program.

# A  Summary of renormalisation scales obtained

| | | $Q^2 = 9\ GeV^2$ | | | $Q^2 = 90\ GeV^2$ | | | $Q^2 = 900\ GeV^2$ | | | $Q^2 = 9000\ GeV^2$ | | |
|---|---|---|---|---|---|---|---|---|---|---|---|---|---|
| $x$ | $y_{cut}$ | $\rho^2_{BLM}$ | $\rho^2_{FAC}$ | $\rho^2_{PMS}$ | $\rho^2_{BLM}$ | $\rho^2_{FAC}$ | $\rho^2_{PMS}$ | $\rho^2_{BLM}$ | $\rho^2_{FAC}$ | $\rho^2_{PMS}$ | $\rho^2_{BLM}$ | $\rho^2_{FAC}$ | $\rho^2_{PMS}$ |
| .0001 | .005 | 1.1 | 0.42 | 0.30 | – | – | – | – | – | – | – | – | – |
| .0001 | .010 | 1.4 | 0.85 | 0.61 | – | – | – | – | – | – | – | – | – |
| .0001 | .020 | 1.9 | 5.0 | 3.6 | – | – | – | – | – | – | – | – | – |
| .0001 | .040 | 3.3 | $1.1 \cdot 10^2$ | 80. | – | – | – | – | – | – | – | – | – |
| .0001 | .080 | 7.5 | $9.9 \cdot 10^3$ | $7.0 \cdot 10^3$ | – | – | – | – | – | – | – | – | – |
| .0003 | .005 | 0.91 | 0.71 | 0.51 | – | – | – | – | – | – | – | – | – |
| .0003 | .010 | 1.1 | 0.69 | 0.50 | – | – | – | – | – | – | – | – | – |
| .0003 | .020 | 1.4 | 2.2 | 1.6 | – | – | – | – | – | – | – | – | – |
| .0003 | .040 | 2.2 | 26. | 18. | – | – | – | – | – | – | – | – | – |
| .0003 | .080 | 4.5 | $1.1 \cdot 10^3$ | $7.6 \cdot 10^2$ | – | – | – | – | – | – | – | – | – |
| .0010 | .005 | 0.73 | 1.9 | 1.4 | 0.67 | 2.3 | 1.7 | – | – | – | – | – | – |
| .0010 | .010 | 0.83 | 0.81 | 0.58 | 0.82 | 0.96 | 0.71 | – | – | – | – | – | – |
| .0010 | .020 | 1.0 | 1.0 | 0.74 | 1.0 | 1.4 | 1.0 | – | – | – | – | – | – |
| .0010 | .040 | 1.5 | 5.2 | 3.7 | 1.4 | 7.0 | 5.2 | – | – | – | – | – | – |
| .0010 | .080 | 2.6 | $1.0 \cdot 10^2$ | 73. | 2.6 | $1.5 \cdot 10^2$ | $1.1 \cdot 10^2$ | – | – | – | – | – | – |
| .0030 | .005 | 0.52 | 7.1 | 5.0 | 0.45 | 6.7 | 4.9 | – | – | – | – | – | – |
| .0030 | .010 | 0.59 | 1.3 | 0.92 | 0.60 | 1.4 | 1.0 | – | – | – | – | – | – |
| .0030 | .020 | 0.75 | 0.71 | 0.51 | 0.75 | 0.68 | 0.50 | – | – | – | – | – | – |
| .0030 | .040 | 1.0 | 1.6 | 1.1 | 0.99 | 1.5 | 1.1 | – | – | – | – | – | – |
| .0030 | .080 | 1.5 | 13. | 9.2 | 1.5 | 15. | 11. | – | – | – | – | – | – |
| .0100 | .005 | 0.25 | 25. | 18. | 0.18 | 21. | 15. | 0.13 | 9.3 | 6.8 | – | – | – |
| .0100 | .010 | 0.30 | 2.9 | 2.1 | 0.30 | 2.5 | 1.8 | 0.24 | 1.5 | 1.1 | – | – | – |
| .0100 | .020 | 0.44 | 0.83 | 0.59 | 0.45 | 0.63 | 0.46 | 0.38 | 0.56 | 0.41 | – | – | – |
| .0100 | .040 | 0.61 | 0.67 | 0.48 | 0.61 | 0.48 | 0.35 | 0.56 | 0.58 | 0.43 | – | – | – |
| .0100 | .080 | 0.87 | 1.7 | 1.2 | 0.86 | 1.5 | 1.1 | 0.83 | 2.1 | 1.5 | – | – | – |
| .0300 | .005 | 0.19 | 2.4 | 1.8 | 0.041 | 41. | 30. | 0.035 | 23. | 17. | – | – | – |
| .0300 | .010 | 0.12 | 2.1 | 1.5 | 0.089 | 3.8 | 2.8 | 0.079 | 2.8 | 2.1 | – | – | – |
| .0300 | .020 | 0.17 | 0.97 | 0.70 | 0.17 | 0.74 | 0.55 | 0.15 | 0.53 | 0.39 | – | – | – |
| .0300 | .040 | 0.29 | 0.42 | 0.30 | 0.29 | 0.25 | 0.18 | 0.27 | 0.24 | 0.18 | – | – | – |
| .0300 | .080 | 0.45 | 0.40 | 0.29 | 0.44 | 0.35 | 0.26 | 0.41 | 0.32 | 0.23 | – | – | – |
| .1000 | .005 | – | – | – | 0.0044 | 42. | 31. | 0.0039 | 29. | 21. | $3.1 \cdot 10^{-3}$ | 21. | 15. |
| .1000 | .010 | – | – | – | 0.011 | 3.4 | 2.5 | 0.097 | 2.5 | 1.8 | $7.6 \cdot 10^{-3}$ | 2.0 | 1.4 |
| .1000 | .020 | – | – | – | 0.026 | 0.52 | 0.38 | 0.024 | 0.41 | 0.30 | $1.8 \cdot 10^{-2}$ | 0.35 | 0.26 |
| .1000 | .040 | – | – | – | 0.061 | 0.15 | 0.11 | 0.054 | 0.13 | 0.092 | $4.1 \cdot 10^{-2}$ | 0.12 | $9.2 \cdot 10^{-2}$ |
| .1000 | .080 | – | – | – | 0.13 | 0.089 | 0.066 | 0.12 | 0.085 | 0.062 | $9.1 \cdot 10^{-2}$ | 0.11 | $8.2 \cdot 10^{-2}$ |
| .3000 | .005 | – | – | – | – | – | – | 0.00057 | 19. | 14. | $5.4 \cdot 10^{-4}$ | 15. | 11. |
| .3000 | .010 | – | – | – | – | – | – | 0.0014 | 1.3 | 0.96 | $1.3 \cdot 10^{-3}$ | 1.1 | 0.83 |
| .3000 | .020 | – | – | – | – | – | – | 0.0032 | 0.16 | 0.12 | $3.1 \cdot 10^{-3}$ | 0.14 | 0.10 |
| .3000 | .040 | – | – | – | – | – | – | 0.0079 | 0.038 | 0.028 | $7.6 \cdot 10^{-3}$ | $3.5 \cdot 10^{-2}$ | $2.6 \cdot 10^{-2}$ |
| .3000 | .080 | – | – | – | – | – | – | 0.019 | 0.014 | 0.010 | $1.8 \cdot 10^{-2}$ | $1.3 \cdot 10^{-2}$ | $9.6 \cdot 10^{-3}$ |
| .9000 | .005 | – | – | – | – | – | – | – | – | – | $2.5 \cdot 10^{-5}$ | $2.0 \cdot 10^{-2}$ | $1.5 \cdot 10^{-2}$ |
| .9000 | .010 | – | – | – | – | – | – | – | – | – | $5.8 \cdot 10^{-5}$ | $5.6 \cdot 10^{-4}$ | $4.2 \cdot 10^{-4}$ |
| .9000 | .020 | – | – | – | – | – | – | – | – | – | $1.3 \cdot 10^{-4}$ | $3.2 \cdot 10^{-5}$ | $1.8 \cdot 10^{-5}$ |
| .9000 | .040 | – | – | – | – | – | – | – | – | – | $3.1 \cdot 10^{-4}$ | $2.7 \cdot 10^{-6}$ | $1.5 \cdot 10^{-6}$ |
| .9000 | .080 | – | – | – | – | – | – | – | – | – | $7.5 \cdot 10^{-4}$ | $3.7 \cdot 10^{-7}$ | $2.0 \cdot 10^{-7}$ |

Table 3: *Obtained renormalisation scales, $\rho^2 = \mu_R^2/Q^2$.*